\documentclass[11pt]{article}
\usepackage{fullpage,citesort,epsfig,psfrag}
\usepackage[normal,small]{caption}

\def\Tr{{\rm Tr}}
\def\to{\rightarrow}
\newcommand{\be}{\begin{equation}}
\newcommand{\ee}{\end{equation}}
\newcommand{\bq}{\begin{eqnarray}}
\newcommand{\eq}{\end{eqnarray}}
\newcommand{\ket}[1]{|#1\rangle}
\newcommand{\bra}[1]{\langle#1|}

\def\ie{{\it i.e.\ }}
\def\m@th{\mathsurround=0pt }
\def\leftrightarrowfill{$\m@th \mathord\leftarrow \mkern-6mu \cleaders\hbox{$\mkern-2mu \mathord- \mkern-2mu$}\hfill
 \mkern-6mu \mathord\rightarrow$}
\def\overleftrightarrow#1{\vbox{\ialign{##\crcr
     \leftrightarrowfill\crcr\noalign{\kern-1pt\nointerlineskip}
     $\hfil\displaystyle{#1}\hfil$\crcr}}}

\begin{document}
\setlength{\captionmargin}{20pt}

\renewcommand{\thefootnote}{\fnsymbol{footnote}}
\begin{titlepage}
\begin{flushright}
LBNL-49078\\
UCB-PTH-01/40\\
UFIFT-HEP-01-20\\
hep-th/0110301
\end{flushright}

\vskip 2.5cm

\begin{center}
\begin{Large}
{\bf A Worldsheet Description of Large $N_c$ Quantum Field
Theory\footnote{This work was supported in part by the Department
of Energy under Grants No. DE-FG02-97ER-41029 and DE-AC03-76SF00098, and in part by tne National Science Foundation
Grant PHY-0098840.}
}
\end{Large}

\vskip 2.cm

{\large 
 Korkut Bardakci\footnote{E-mail  address: {\tt kbardakci@lbl.gov}}
and Charles B. Thorn\footnote{Visiting Miller Research Professor,
on sabbatical leave from the Department of
Physics, University of Florida, Gainesville, FL 32611.}\footnote{E-mail  address: {\tt thorn@phys.ufl.edu}}}
\vskip 0.4cm

{\it Department of Physics, University of California,
Berkeley CA 94710
}
\vskip0.15cm
and
\vskip0.15cm
{\it Theoretical Physics Group, Lawrence Berkeley National
Laboratory\\ University of California,
Berkeley CA 94710
}

\vskip 1.0cm
\end{center}

\begin{abstract}\noindent
The $N_c\to\infty$ limit of a matrix quantum field theory is
equivalent to summing only planar Feynman diagrams. The
possibility of interpreting this sum as some kind
world-sheet theory has been in the air ever since
't Hooft's original paper. We establish here just such
a world sheet description for a scalar quantum
field with interaction term $g\Tr\phi^3/\sqrt{N_c}$, and we
indicate how the approach might be extended to more
general field theories. 
\end{abstract}
\vfill
\end{titlepage}
\section{Introduction}
\label{chap1}
Almost from the beginning of the development of string theory
there has been the suspicion that underlying it is a
local quantum field theory. The first concrete proposal
of such a connection was the fishnet diagram model of
Nielsen and Olesen and Sakita and Virasoro \cite{nielsenfishnet}.
These authors suggested identifying the string worldsheet with
very large planar diagrams. 

Then 't Hooft showed how
to single out the planar diagrams of an $N_c\times N_c$
matrix quantum field theory by taking the limit $N_c\to\infty$
\cite{thooftlargen}. However, the large fishnet
diagrams are only a tiny subset of the surviving planar
diagrams in this limit, and so it was not at all clear
how to interpret the sum of them all as a string worldsheet.
Thus there was the suggestion that strong 't Hooft
coupling $\lambda=N_c g^2$ was really needed to
force the dominance of the large fishnet diagrams
with a consequent worldsheet interpretation of that
double limit \cite{thornfishnet,bardakcis}. 
Another version of this idea was given in \cite{brezink,douglass,grossm}
where large planar diagrams were enhanced by going to
a critical 't Hooft coupling as well as large $N_c$. 
The need for this double limit has also arisen in the
more modern version of the string/QFT connection as
proposed by Maldacena \cite{maldacena,gubserkp,wittenholog}.
Other recent work on worldsheets from large planar diagrams
can be found in \cite{siegelfish,beringrt}.

But, in addition to his association of planar diagrams
with the $N_c\to\infty$ limit, 't Hooft also hinted at
a way to associate a smooth world sheet with {\it each}
planar diagram, large or small, at least in light-cone
gauge \cite{thooftlargen}. If such a formulation
is possible, the sum of the planar diagrams of
a quantum field theory can be thought of in the same
way as the sum of interacting open string 
diagrams \cite{mandelstam,gilest}. We think such a
formulation of the sum of all planar diagrams 
would be an important step toward understanding
the string/QFT connection. For example, if properly
formulated one might be able to identify a worldsheet order
parameter whose expectation would be the string tension.
One might then hope to learn a lot about the
physics of the emergent string by following the
consequences of such a non-vanishing order
parameter, even if
a complete solution of the dynamics is not available.

In this article
we give such an interpretation for the particularly simple
case of the planar diagrams of a massless scalar field theory
with cubic interactions. We are not concerned here about the
intrinsic instability of this theory, because our
construction is a reinterpretation of each individual
Feynman diagram in terms of a world sheet, \ie it
works order by order in perturbation theory. Thus
our real aim is to provide a paradigm for a worldsheet 
description of the planar diagrams of a wide
range of quantum field theories.

We shall use light-cone quantization in our initial
construction, although we shall see that the
resulting light-cone worldsheet model can be 
extended to a covariant description. In the context
of the light-cone description, one can understand
the physics underlying our approach by considering
the infinite rest tension limit of string theory,
which is supposed to be a quantum field theory. 
The planar multi-loop diagrams of the string theory
should then go over to the planar multi-loop
diagrams of the limiting field theory. But the
light-cone string treats $p^+$ in a very special
way. The parameter $\sigma$  marking points on the string
is chosen so that the $p^+$ density is unity so $p^+$ is just the 
length of the parameter interval $0\leq\sigma\leq p^+$. To
think more clearly about what this means, it
is helpful to discretize $\sigma=lm$, $l=1,\ldots, M$ 
and hence $p^+=Mm$. Then the string is a bound composite 
system of string bits, each carrying one unit
$p^+$. If one takes the $T_0\to\infty$ limit
of the light-cone multi-loop diagrams discretized
in this way, the string
is indeed tightly bound in transverse space, but
the $p^+$ it carries is still uniformly distributed
among the bits \cite{thornfront}. From this it
is clear what we need to do. We must think
of a field quantum carrying $p^+=Mm$, not
as a single particle with $M$ units of $p^+$,
but rather as $M$ bits, each with one unit of $p^+$.

But this entails introducing transverse coordinates and
momenta for each bit. Our construction places these
bit degrees of freedom in the bulk of the worldsheet.
All but one of these degrees
of freedom must be redundant, so we expect that
the worldsheet we construct will be ``topological''
in the sense that all the physics must live on the
boundaries. Thus our work has parallels with
other studies of topological string theories,
see for example \cite{horavatopqcd}.

In Section 2 we begin our work by constructing
a local world sheet representation of the
free scalar propagator on the light-cone.
In Section 3 we show how the light-cone worldsheet
action we infer can be extended to a covariant
one. In Section 4 we extend the construction to
include the cubic interactions of $\Tr\phi^3$
matrix field theory and propose
a worldsheet path integral which sums all
of the planar loop diagrams by coupling the worldsheet
variables to an Ising spin system set up on the sites
of the world sheet. In Section 5 we give a
representation of this Ising spin system in terms
of Grassmann integration. This new formulation
allows the continuum limit of the coupled Ising
system to be taken at least formally. In Section 6
we delve into the issue of giving the fields a mass.
In a short section 7 we indulge in some amusing 
comments on duality between string worldsheets and
QFT worldsheets. Finally, in Section 8 we give
a list of the many generalizations and 
potential developments we leave for the future. 
\section{Free Massless Scalar Field}
In this section we shall use light-cone coordinates defined
as $x^\pm=(x^0\pm x^3)/\sqrt2$. The remaining components
of $x^\mu$ will be distinguished by Latin indices, or
as a vector by bold-face type. Thus the four coordinates
$x^\mu$ will be $(x^+,x^-,{\bf x})$ or $(x^+,x^-,x^k)$.
We shall similarly use the same conventions for the
components of any four vector. Thus the Lorentz invariant
scalar product of two four vectors $v,w$ is written
$v\cdot w={\bf v}\cdot{\bf w}-v^+w^--v^-w^+$. We shall
select $x^+$ to be our quantum evolution parameter, and
recall that the ``energy'' conjugate to this time is
$p^-$. A massless on-shell particle thus has the ``energy''
$p^-={\bf p}^2/2p^+$.

We begin our discussion 
with the mixed representation of the propagator of free massless
matrix scalar field \cite{thooftlargen}:
\begin{eqnarray}
\langle \phi_\alpha^\beta(x^+)\phi_\gamma^\delta(0)\rangle
\equiv\delta_\alpha^\delta\delta_\gamma^\beta\int {dp^-\over2\pi i}e^{-ix^+p^-}{1\over p^2-i\epsilon}
=\delta_\alpha^\delta\delta_\gamma^\beta{\theta(x^+)\over2p^+}e^{-ix^+{\bf p}^2/2p^+}\to\delta_\alpha^\delta\delta_\gamma^\beta
{\theta(\tau)\over2p^+}e^{-\tau{\bf p}^2/2p^+},
\label{propagator}
\end{eqnarray}
where we assume $p^+>0$. Since we shall consider only planar
diagrams in this paper, we shall suppress the
color factors $\delta_\alpha^\delta\delta_\gamma^\beta$
in the rest of the paper. Note that we have defined imaginary
time $\tau=ix^+$. We use imaginary $x^+$ 
for convenience to make our integrals damped
instead of rapidly oscillating. It is certainly
not essential for our purposes.
For instance, we shall 
return to real time when discussing
the classical limit. The expression (\ref{propagator}) 
is the starting point
for our worldsheet construction. 

Based on the analogy
of the light-cone string \cite{goddardgrt}, 't Hooft
associated this propagator with a rectangular world
sheet of width $p^+$ and length $\tau$. The suppressed
color factors are associated with spatial boundaries of
this rectangle. We shall
do the same. The expression for the propagator, however,
is  not yet associated in any way with local variables
on this world sheet.

To facilitate the introduction of such local variables
we set up a lattice worldsheet by discretizing $\tau$
and $p^+$:
\begin{eqnarray}
\tau= ka,\qquad p^+=lm,\qquad {\rm for}\quad k,l=1,2,3,\ldots
\end{eqnarray}
Then the scalar propagator becomes
\begin{eqnarray}
{\theta(k)\over2lm}e^{-k(a/m){\bf p}^2/2l}
\end{eqnarray}
The factor $1/2lm$ is better associated with one of the
two vertices to which the line is attached. Here we have a
choice between assigning the factors symmetrically
by using square roots, or asymmetrically. The asymmetry
is natural in light-cone parameterization because
all propagation is forward in time. Thus we can,
for example, assign the factor to the {\it earlier}
of the two vertices connected by the line. Then a {\it fission}
vertex, in which one particle with $p^+=Mm$ ``decays'' to two particles
with $p^+_1=M_1m$, $p^+_2=M_2m$ will be assigned the
factor $1/ 4M_1M_2m^2$, whereas a fusion vertex, which
is the time-reversal of the fission vertex will be assigned
the factor $1/2Mm$. After assigning the prefactors in this
way, all propagators are then simple exponentials $e^{-ka{\bf p}^2/2lm}$.
 
Consider a line carrying $M$ units of $p^+$. As in \cite{thooftlargen}
it will be
convenient to write the total momentum as a difference  
${\bf p}\equiv {\bf q}_M-{\bf q}_0$. For each discrete
time step, we notice that one can write
\begin{eqnarray}
\exp\left\{-{a\over m}
{({\bf q}_M-{\bf q}_0)^2\over2M}\right\}=
M^{d/2}\left({a\over2\pi m}\right)^{d(M-1)/2}\int d{\bf q}_1
d{\bf q}_2\cdots d{\bf q}_{M-1}e^{
-{a}\sum_{i=0}^{M-1}({\bf q}_{i+1}-{\bf q}_i)^2/2m}.
\label{bosebits}
\end{eqnarray}
We would like to use this identity to interpret a line carrying
$M$ units of $p^+$ as $M$ lines, each carrying a single
unit of $p^+$. At first sight it seems that the factors
multiplying the integral obstruct this interpretation. 
Indeed, there is an essential difference between a single
particle state and a multi-particle state, in that the
latter has many momenta assigned to it, \ie the phase space is
much larger. The extra particles must be ``fictitious'' in
some appropriate sense. Alternatively, there must be some 
strong binding force that binds the many particles together
to act as a single particle \cite{thornfront}. In this
article we choose the former possibility.

To achieve this ``fictitiousness'', we introduce a pair of
anti-commuting ghost fields
$b_i, c_i$ for every two transverse dimensions. Consider then
the ghost integrals
\begin{eqnarray}
\int \prod_{i=1}^{M-1} dc_idb_i \exp\left\{
\sum_{i=0}^{M-1}(b_{i+1}-b_i)(c_{i+1}-c_i)\right\}
&=&Me^{(b_{M}-b_0)(c_{M}-c_0)/M}\nonumber\\
&=& M \qquad{\rm for}\quad c_0=c_{M}=0, b_0=b_{M}=0.
\label{ghostbits}
\end{eqnarray}
To get the required factor of $M$ we adopt the Dirichlet conditions
indicated.
For simplicity of presentation, let us restrict our attention to
the case $d=2$, which is actually the case of most interest
(four dimensional space-time). For this case we
only require one $b,c$ set of ghosts. Then, we may rewrite the identity
of Eq~(\ref{bosebits})
\begin{eqnarray}
\exp\left\{-{a\over m}
{({\bf q}_M-{\bf q}_0)^2\over2M}\right\}&=&
\nonumber\\
&&\hskip-3cm\int\prod_{i=1}^{M-1} {dc_idb_i\over2\pi} d{\bf q}_i\exp\left\{{a\over m}\sum_{j=0}^{M-1}(b_{j+1}-b_j)(c_{j+1}-c_j)-
{a\over2m}\sum_{i=0}^{M-1}({\bf q}_{i+1}-{\bf q}_i)^2\right\},
\label{boseghostbits}
\end{eqnarray}
where we recall that we have imposed $c_0=b_0=c_M=b_M=0$.
To build the propagator for $N$ time steps, we merely repeat this
construction $N$ times:
\begin{eqnarray}
\exp\left\{-N{a\over m}
{({\bf q}_M-{\bf q}_0)^2\over2M}\right\}&=&
\nonumber\\
&&\hskip-5cm\int\prod_{j=1}^N\prod_{i=1}^{M-1} {dc^j_idb^j_i\over2\pi} 
d{\bf q}^j_i\exp\left\{{a\over m}\sum_j\sum_{i=0}^{M-1}
(b^j_{i+1}-b^j_i)(c^j_{i+1}-c^j_i)-
{a\over2m}\sum_j\sum_{i=0}^{M-1}({\bf q}^j_{i+1}-{\bf q}^j_i)^2\right\}.
\label{nstepbits}
\end{eqnarray}

The expression (\ref{nstepbits}) implements momentum conservation
essentially ``by hand''. The momentum propagated ${\bf p}={\bf q}_M-{\bf q}_0$
is conserved by the imposition of the Dirichlet boundary conditions
${\bf q}_{0,j}={\bf q}_0$ and ${\bf q}_{M,j}={\bf q}_M$. To 
incorporate these conditions in the world sheet path integral
requires the insertion of delta functions that conserve these
boundary values from one time slice to the next. This will
be a crucial refinement for the description of multi-loop
diagrams. The most economical approach is to retain the
Dirichlet boundary condition at one end of the strip, say $i=0$, where
we impose ${\bf q}_0^j={\bf q}_0$ but insert
momentum conserving delta functions at the other end $i=M$.
It is convenient but not necessary to set ${\bf q}_0=0$.
In fact the particle transition amplitude {\it should} have these
delta functions explicitly incorporated:
\begin{eqnarray}
T_{fi}=\bra{{\bf p}_f}e^{-\tau p^-}\ket{{\bf p}}&=&\delta({\bf p}_f-{\bf p})
e^{-\tau{\bf p}^2/2p^+}\\
&\to&\int\prod_{j=1}^N d{\bf q}_M^j\prod_{j=0}^N
\delta({\bf q}_M^{j+1}-{\bf q}_M^j)e^{-a\sum_j({\bf q}_M^j-{\bf q}_0)^2/2p^+}
\nonumber\\
&=&\int\prod_{j=1}^N d{\bf q}_M^j\prod_{j=0}^N{d{\bf x}_M^j\over(2\pi)^d}
e^{-i{\bf x}_M^j\cdot({\bf q}_M^{j+1}-{\bf q}_M^j)}
e^{-a\sum_j({\bf q}_M^j-{\bf q}_0)^2/2mM}.
\end{eqnarray}
The complete path integral is then reached by inserting (\ref{nstepbits})
for $d=2$.
\begin{eqnarray}
T_{fi}&=&
\int\prod_{j=1}^N\prod_{i=1}^{M} {dc^j_idb^j_i\over2\pi} 
d{\bf q}^j_i\prod_{j=0}^N{d{\bf x}_M^j\over(2\pi)^2}
\exp\left\{-i\sum_{j=0}^N{\bf x}_M^j\cdot({\bf q}_M^{j+1}-{\bf q}_M^j)\right\}\nonumber\\
&&\hskip0cm\exp\left\{
{a\over m}\sum_j\sum_{i=0}^{M-1}
(b^j_{i+1}-b^j_i)(c^j_{i+1}-c^j_i)-
{a\over2m}\sum_j\sum_{i=0}^{M-1}({\bf q}^j_{i+1}-{\bf q}^j_i)^2\right\}.
\label{discretefreepi}
\end{eqnarray}

This completes the worldsheet construction for the propagator for
a free scalar field. It is represented by a strip of length $T=(N+1)a$
and width $p^+=Mm$. The discretized world-sheet is a rectangular 
grid. Each site carries a momentum variable ${\bf q}_i^j$ and
a pair of ghost fields $b_i^j,c_i^j$. In addition the sites at
the $i=M$ boundary carry coordinates ${\bf x}_{M}^j$.
The discretized world sheet
action is just the exponent appearing in (\ref{discretefreepi}).
The total transverse momentum carried by the propagator is
the difference of the momenta, ${\bf p}={\bf q}_M-{\bf q}_0$, 
at the boundaries of the strip, which are fixed by the 
Dirichlet conditions imposed by the delta functions. 
Notice that there are no terms
in the bulk action coupling different time slices: the free world
sheet dynamics is ``constrained'' or topological. For the
free propagator time slice couplings occur only at the
boundaries. When
we draw discretized worldsheet diagrams (see Fig.~\ref{oneloopws}),
with time flowing up,
we shall distinguish the bulk and boundary degrees of freedom
by dotted and solid vertical lines respectively. As we shall see, the field
theoretic interactions will gradually introduce more
time slice couplings, 
via the insertion of new Dirichlet boundaries (indicated by solid lines) 
within the worldsheet, order by order in perturbation theory. 

In the next section we discuss the continuum path integral in a more general
parameterization than light-cone, so that we can expose the
Lorentz covariance of the description more clearly. For this
purpose it is convenient to write the boundary terms involving
the coordinates as a topological bulk expression. We remark here
that we can do this even at the discretized level by noting the
equality
\begin{eqnarray}
\sum_{j=0}^N{\bf x}_M^j\cdot({\bf q}_M^{j+1}-{\bf q}_M^j)&=&
\sum_{j=1}^N\sum_{i=0}^{M-1}\left[({\bf x}_{i+1}^j-{\bf x}_{i}^j)\cdot
({\bf q}_i^{j+1}-{\bf q}_i^j)-({\bf x}_{i+1}^j-{\bf x}_{i+1}^{j-1})\cdot
({\bf q}_{i+1}^{j}-{\bf q}_i^j)\right]\nonumber\\
&&\hskip2cm + S_N-S^\prime_0,
\end{eqnarray}
where the last two terms only involve variables at final and initial
times $j=N,N+1$ and $j=0,1$ respectively. They therefore do not
affect the dynamics described by the free field propagator.
\newpage
\section{Continuum Limit and Lorentz Invariance}
\subsection{Free Propagator}
In this section we give a first discussion
of the continuum limit of the world sheet construction for the
free propagator. We note that the
worldsheet path integral has the formal continuum limit
\begin{eqnarray}
T_{fi}&=&
\int DcDbD{\bf q}D{\bf x}
\exp\left\{i\int d\tau d\sigma \left({\bf \dot x}\cdot{\bf q}^\prime
-{\bf x}^\prime\cdot{\bf \dot q}\right)\right\} \prod_{\sigma,\tau}
\delta({\bf x}^\prime(\sigma,\tau))
\nonumber\\&&
\exp\left\{\int_0^T d\tau\int_{0}^{p^+}d\sigma
b^\prime(\sigma,\tau)c^\prime(\sigma,\tau)-
{1\over2}\int_0^T d\tau\int_{0}^{p^+}d\sigma{\bf q}^\prime(\sigma,\tau)^2
\right\}.
\label{naivecont}
\end{eqnarray}
To interpret this formula more covariantly, we first of
all return to real time, $\tau\to i\tau$, and extend the
transverse components of momentum  $q^k(\sigma,\tau)$ to
a four vector $q^\mu(\sigma,\tau)$ by introducing $q^\pm(\sigma,\tau)$.
Then we recognize that we have actually fixed the $\sigma$ parameter
so that $q^+(\sigma,\tau)=\sigma$ or $q^{+\prime}(\sigma,\tau)=1$.
Then we have 
$$q^\prime\cdot q^\prime={\bf q^\prime}^2
-2q^{+\prime}q^{-\prime}={\bf q^\prime}^2
-2q^{-\prime}.$$
The last term in the second exponent of (\ref{naivecont})
just represents the energy contribution to the action
$-\int d\tau p^-$. With our identification of momentum
components as differences, $p^-=q_1^--q^-_0=\int d\sigma q^{-\prime}$,
we see that this identification follows in this
parameterization from the covariant constraint
$q^\prime\cdot q^\prime = 0$. This suggests that we
start in a more general parametrization with the action
\begin{eqnarray}
S=\int d\tau \int_{\sigma_0(\tau)}^{\sigma_1(\tau)} d\sigma 
\left({\dot x}\cdot{q^\prime}-{x^\prime}\cdot{\dot q}
-{\lambda\over2}q^{\prime2}
\right)
\label{covact}
\end{eqnarray}
where $\lambda$ is a Lagrange multiplier enforcing the
constraint $q^{\prime2}=0$, and we recall that the value of $q^\mu$
is fixed at $\sigma=\sigma_0(\tau)$. 

We notice the invariances of (\ref{covact}). There is invariance
under a partial reparametrization under
\begin{eqnarray}
\sigma\to \sigma(\tau^\prime,\sigma^\prime),\qquad \tau\to\tau(\tau^\prime).
\label{partdiffeo}
\end{eqnarray} 
In addition, due to the topological
nature of the first term in the action, there is a bulk gauge invariance
\begin{eqnarray}
x^\mu(\sigma,\tau)\to x^\mu(\sigma,\tau) + \Lambda^\mu(\sigma,\tau),\qquad
{\rm with}\quad \Lambda(\sigma_1,\tau)=0. 
\label{gaugesym}
\end{eqnarray}
The restriction to $\Lambda(\sigma_1)=0$ is necessary because of
the boundary value of ${\bf x}$ appears explicitly in the action.

We would like to point out that it is possible to start with an action
invariant under an even larger group, namely,
unrestricted reparametrizations:
\be
\sigma\rightarrow \sigma(\tau',\sigma'), \;\; \tau\rightarrow
\tau(\tau',\sigma').
\ee
This is achieved by introducing the usual world sheet metric 
$g^{\alpha \beta}$ and writing
\be
S=\int d\tau \int d\sigma ({\dot x}\cdot q^\prime-x^\prime\cdot{\dot q}
+g^{\alpha \beta} \partial_{\alpha} q
\cdot \partial_{\beta} q),
\ee
where $\alpha=0$ refers to $\tau$ and $\alpha=1$ to $\sigma$. In
addition, we impose the crucial constraint
\be
\det(g)=0.
\ee
This constraint distinguishes our action from the usual string
action. There is no conformal gauge available anymore; instead,
one can gauge fix partially by setting all the components of
$g^{\alpha \beta}$, with the exception of $g^{1 1}$ equal to
zero. Identifying $g^{1 1}$ with $-\lambda/2$, the form of the
action given by Eq~(\ref{covact}) is then recovered, with its residual 
reparametrization invariance. We will not pursue this more
general formulation further in this paper, however, it may
provide a better starting point for possible future investigations.

As a check of our proposed action, we
now show how to regain the completely gauge fixed action
from (\ref{covact}) by exploiting the
local symmetries (\ref{partdiffeo},\ref{gaugesym}). 
First, we fix $\sigma$ reparametrization invariance
by choosing $\sigma=q^+$.  
Then the $x^{-\prime}$ term drops out of the action and the
equation of motion for $x^-$ is simply $p^+$
conservation. As in the light-cone string this
condition is  conveniently
implemented by restricting the $\sigma$ parameter range to
$\tau$ independent boundaries: $\sigma_0=q^+_0$, $\sigma_1=q^+_1$. 
After these choices the action reduces to
\begin{eqnarray}
S=\int d\tau \int_{q^+_0}^{q^+_1} d\sigma 
\left({\bf\dot x}\cdot{{\bf q}^\prime}-{\dot x}^+q^{-\prime}
-{\bf x}^\prime\cdot{{\bf\dot q}}+x^{+\prime}{\dot q}^-
-{\lambda\over2}\left({\bf q}^{\prime2} - 2q^{-\prime}\right)\right)
\label{fixedact2}
\end{eqnarray}
Integrating out $q^-$  then
gives $\lambda^\prime=0$ and $\lambda(\sigma_1)
={\dot x}^{+}(\sigma_1)$, leading to
\begin{eqnarray}
S=\int d\tau \int_{q^+_0}^{q^+_1} d\sigma 
\left({\bf\dot x}\cdot{{\bf q}^\prime}
-{\bf x}^\prime\cdot{{\bf\dot q}}
-{{\dot x}^+(\sigma_1,\tau)\over2}\ {\bf q}^{\prime2}\right)
\label{fixedact3}
\end{eqnarray}
The gauge invariance (\ref{gaugesym})
can now be exploited to set ${\bf x}^\prime=0$,
as a constraint in the path integral. At the same
time we are also free to set $x^{+\prime}=0$, since
only $x^+(\sigma_1,\tau)$ remains in the action.
With $x^+$ independent of $\sigma$, 
we use the $\sigma$ independent $\tau$ reparametrization invariance
to choose $x^+=\tau$, giving finally
\begin{eqnarray}
S=\int d\tau \left[{\bf\dot x}\cdot({\bf q}_1(\tau)
-{\bf q}_0)-{1\over2}\int_{0}^{p^+}d\sigma\ {\bf q}^{\prime2}\right]
\to \int d\tau \left[-{\bf x}\cdot{\bf \dot q}_1(\tau)
-{1\over2}\int_{0}^{p^+}d\sigma\ {\bf q}^{\prime2}\right]
\label{fixedact1}
\end{eqnarray}
as desired.

Finally, we discuss Fadeev-Popov ghosts. Since we
are able to completely eliminate the $\pm$ components of
$x,q$ by our parametrization choice, we have also implicitly eliminated 
the corresponding ghosts. However, the bulk transverse components remain, and
we should retain those gauge fixing ghosts. Since the
gauge condition on ${\bf x}$ was ${\bf x}^\prime=0$, the corresponding
F-P determinant is 
$$({\det}_{\rm ND}(i\partial/\partial\sigma))^{d}
=({\det}_{\rm ND}(-\partial^2/\partial\sigma^2))^{d/2}=1.$$ 
Here the subscript $\rm ND$ indicates mixed Neumann-Dirichlet
boundary conditions due to the restrictions (see (\ref{gaugesym})
on the gauge parameter for $x$. To see that this determinant
is unity, put $c_0=b_0=0$
and integrate over $c_M,b_M$ in the first of
Eqs.~(\ref{ghostbits}).

In contrast, the Parisi-Sourlas ghost factor in (\ref{naivecont})
is  ${\det}_{DD}(-\partial^2/\partial\sigma^2)$ with 
Dirichlet conditions at {\it both} ends. We can identify
this factor as a Fadeev-Popov determinant if we interpret
the ${\bf q}^{\prime2}$ term in (\ref{naivecont}) as a
Feynman-style gauge fixing term for the ${\bf q}$ gauge invariance
\begin{eqnarray}
{\bf q}(\sigma,\tau)\to{\bf q}(\sigma,\tau)
+{\bf \Lambda}^\prime(\sigma,\tau),
\qquad {\rm with}\quad{\bf \Lambda}^\prime(\sigma_0,\tau)=
{\bf \Lambda}^\prime(\sigma_1,\tau)=0,
\end{eqnarray}
enjoyed by the topological part of $S$.
Here the restriction on ${\bf \Lambda}^\prime$ is necessary
because ${\bf q}$ satisfies Dirichlet conditions at both
ends. The F-P factor for this gauge fixing is the determinant
of the same differential operator as in the $x$ case, but
this time with Dirichlet conditions at both ends, \ie
precisely the ghost factor shown in (\ref{naivecont}).
\subsection{Interactions}
\begin{figure}[ht]
\vskip1cm
\centerline{\epsfig{file=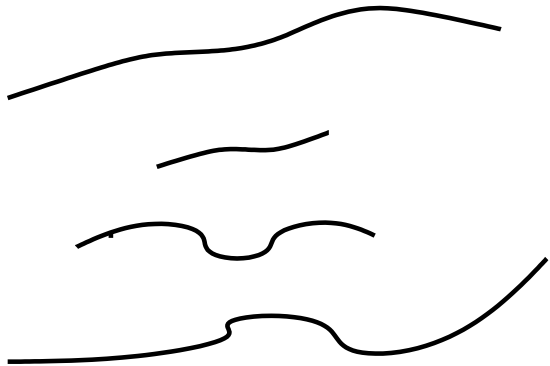,width=8cm}}
\caption{World sheet in a general parameterization: none
of the solid lines need  be straight.}
\label{curvy}
\vskip1cm
\end{figure}

The above discussion dealt exclusively with the free
propagator. When interactions are included as in the 
following section, we must generalize the discussion
of covariance to the 
more general case of a worldsheet with internal
solid lines as well as external boundaries.
Our starting point is Eq.(\ref{covact}), but with the boundaries
unspecified at the moment.
When applied to a free propagator, the variables $x$ and $q$ were
continuous, except possibly at the boundaries. The generalization to the
interacting case is accomplished by allowing $x$ to be discontinuous
along various segments of curves (see Fig.~\ref{curvy}). 
In contrast,
$q$ is restricted to be continuous everywhere.
 The beginning and the end
of the segment is where the interaction takes place, and the functional
integral involves a sum over all possible segments and all possible
discontinuities. This sum is not well defined till we fix the parameters
$\sigma$ and $\tau$, which we will do next.
 Making use of the $\sigma$
reparametrization invariance (\ref{partdiffeo}), we choose $\sigma=q^{+}$,
as above. Since $q^{+}$, which is the same as $p^{+}$ with a
suitable boundary condition, is conserved, the segments of discontinuity
become straight lines, located at various constant values of $\sigma$.
Let the nth segment be at $\sigma=\sigma_{n}$, with the end points
beginning at $\tau=\tau_{n}^{i}$ and ending at $\tau_{n}^{f}$. The
functional integral over the segments then reduces to ordinary integrals
over $\sigma_{n}$, $\tau_{n}^{i}$, and $\tau_{n}^{f}$, plus a sum over
the number of segments. These then correspond to the solid lines of
Fig.(\ref{oneloopws}).

Next we use the gauge invariance (\ref{gaugesym}) to eliminate the variable $x$. It is
important to realize that, in contrast to $x$, the gauge parameter $\Lambda$
is continuous everywhere. Therefore, we can gauge away all of $x$, except
for its discontinuity across a solid line. Or alternatively, since the term
$$
\dot{x}\cdot q'-x'\cdot \dot{q}= \frac{1}{2}\partial_{\tau}(x\cdot q'
-q\cdot x')-
\frac{1}{2}\partial_{\sigma}(x\cdot \dot{q}- q\cdot \dot{x})
$$
is an exact differential, it can be integrated out, leaving behind
boundary contributions from the solid lines, where x is discontinuous.
As a result, S can be rewritten as
\be
S= -\frac{1}{2}\int d\tau \int d\sigma\, \lambda\, q'^{2} +
\sum_{n} \int d\tau\,
 y(\sigma_{n},\tau)\cdot\dot{q}(\sigma_{n},\tau),
\ee
where $y$ denotes the discontinuity of $x$ across a solid line.

 We now
investigate the equations of motion resulting from this action
, with the gauge choice 
$\sigma=q^{+}$. The equation obtained by varying $q^{-}$ gives
\be
\frac{1}{2} \lambda'= \sum_{n}\dot{y}^{+}_{n} \delta(\sigma -\sigma_{n}),
\ee
with $y^{+}_{n}=y^{+}(\sigma_{n},\tau)$. This means that $\lambda$ is a
constant, except for jumps across the solid lines. We can eliminate
these discontinuities with a suitable $\tau$ reparametrization, since
we are allowed independent $\tau$ reparametrization in each
strip bounded by solid lines.
By choosing two different $\tau$ reparametrizations
$\tau\rightarrow f(\tau)$, $\lambda\rightarrow
\lambda f'(\tau)$ across the solid line, the 
discontinuity in $\lambda$ can be eliminated.
As a bonus, this also ensures
\be
\dot{y}^{+}_{n}=0
\label{continuoustime}
\ee
except, of course, at the beginning and the end of the solid lines.
Notice that so far only relative $\tau$ reparametrizations across 
solid lines have been fixed, and an overall reparametrization is
still at our disposal. Now (\ref{continuoustime}) means that $\dot{x}^{+}$ has no jumps
across solid lines and is therefore independent of $\sigma$, and using
the overall $\tau$ reparametrization, we can set $\dot{x}^{+}=1$, or
$x^{+}=\tau$. Also, we obtain the condition $q'^{2}=0$ by varying
with respect to $\lambda$ as in the case of free propagator.

Next, we will study the equations of motion for ${\bf q}$, the transverse
components. These are continuous across the solid lines, and they are
$\tau$ independent, by virtue of the equations resulting from
varying ${\bf y}$. Varying with respect to ${\bf q}$ gives
\be
\lambda\, {\bf q}''= \sum_{n} \dot{{\bf q}}\, \delta(\sigma -\sigma_{n}).
\ee
Therefore, although ${\bf q}$ is continuous, its derivative with respect
to $\sigma$ jumps across a solid line. This equation determines ${\bf q}$
in terms of its values ${\bf q}_{n}$ on the solid lines.
For example, in the region
between $\sigma_{n-1}$ and $\sigma_{n}$, ${\bf q}$ is given by
\be
 {\bf q}=\frac{1}{\sigma_{n}-\sigma_{n-1}} \left({\bf q}_{n}
(\sigma - \sigma_{n-1}) + {\bf q}_{n-1}(\sigma_{n}-\sigma)\right).
\ee
The mass shell condition $q'^{2}=0$ translates into the
equation (recall that $\sigma_k=q^+_k$ by our gauge choice)
\be
q^{-\prime}= \frac{1}{2}\left(\frac{{\bf q}_{n}-{\bf q}_{n-1}}
{q^+_{n}-q^+_{n-1}}\right)^{2} = \frac{1}{2}
\frac{({\bf p}_{n})^{2}}{(p^{+}_{n})^{2}}
\ee
for the $ q^{-}$, when $\sigma$ is between $\sigma_{n-1}$
and $\sigma_{n}$. 
Here we have taken into account
that
$$
{\bf q}_{n}-{\bf q}_{n-1}= {\bf p}_{n},
$$
where ${\bf p}_{n}$ is the momentum flowing through the strip bounded
by the solid lines at $\sigma_{n}$ and $\sigma_{n-1}$.

 This equation can be integrated and 
we get
\be
 p^{-}=q^-_n-q^-_0=\sum_{k=1}^{n} \frac{1}{2} \frac{({\bf p}_{k})^{2}}
{p^{+}_{k}}.
\ee
 Identifying $p^-\equiv q^{-}_{n}-q^-_0$ with the
total energy (conjugate to $\tau=x^{+}$)
flowing through the strip bounded by the solid lines labeled by
$k=0$ and $k=n$, we get the correct result in agreement with the
light cone Feynman rules.

\section{Cubic Interactions: An Ising Spin system}
In this section we show how to include the cubic
vertices in the planar
diagrams (those surviving the $N_c\to\infty$ limit)
of $g\Tr\phi^3/3\sqrt{N_c}$ theory using the discretized
worldsheet formalism set up in Section 2.
Let us first consider a fission vertex, describing a
field quantum with momenta ${\bf Q},Mm$ transforming to
two field quanta with momenta ${\bf p},lm; ({\bf Q}-{\bf p}),(M-l)m$. 
Before the interaction
we have a single propagator treated as in Section 2. After the interaction
we have two propagators. According to our previous discussion,
we must also  assign the extra factor $g/4l(M-l)m^2$ to the vertex
\footnote{Strictly speaking there is also a factor of $1/\sqrt{N_c}$
assigned to each vertex. However within a multiloop planar diagram
there is a color factor $N_c$ associated with each pair
of cubic vertices connecting internal lines, so all the $N_c$
dependence associated with the internal structure of the
diagram cancels, leaving only a factor $(1/\sqrt{N_c})^{E-2}$,
where $E$ is the number of external lines, which we suppress.}.
The challenge is to bring in the factors $1/l(M-l)$ by a
worldsheet local treatment of the fission point.

\begin{figure}[ht]
\vskip1cm
\psfrag{'k'}{$k$}
\psfrag{'k+l'}{$k+l$}
\psfrag{'M'}{$M$}\psfrag{'M1'}{$M_1$}
\psfrag{'N'}{$N$}
\centerline{\epsfig{file=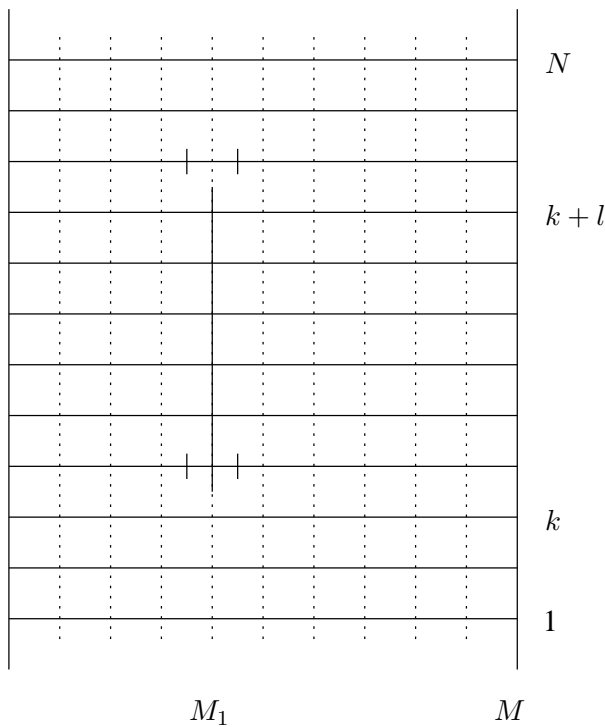,width=8cm}}
\caption{One loop correction to the propagator. The momenta
on the sites crossed by the solid vertical line are all
equal. The tick marks on the four horizontal links just
after the interaction points indicate the deleted ghost
terms.}
\label{oneloopws}
\vskip1cm
\end{figure}

The key to arranging the correct prefactors is to realize
that by simply altering the ghost integral near
the interaction point we can force the ghost integral
to give unity instead of a factor of $M$ or $l(M-l)$.
For example, notice that deleting the $i=M-1$ term in
the exponential of (\ref{ghostbits}) has precisely this effect:
\begin{eqnarray}
\int \prod_{i=1}^{M-1} dc_idb_i \exp\left\{\sum_{i=0}^{M-2}(b_{i+1}-b_i)(c_{i+1}-c_i)\right\}
&=&1.
\label{ghostbitI}
\end{eqnarray}
Note, by the way, that if the first and 
last terms were deleted the result would be
zero:
\begin{eqnarray}
\int \prod_{i=1}^{M-1} dc_idb_i \exp\left\{
\sum_{i=1}^{M-2}(b_{i+1}-b_i)(c_{i+1}-c_i)\right\}
&=&0.
\label{ghostbit0}
\end{eqnarray}
So for the fission vertex, we can choose to delete the corresponding
terms on both sides of the interaction point and for the time
slice just after the fission, and this will
lead to a factor $1/l(M-l)$. By the same token if we delete these 
terms just after a fusion vertex, we will obtain the factor $1/M$,
just as desired. The fission vertex is represented by a time line on
which the momentum is constant. That is, the momentum on this
line is fixed (Dirichlet condition), but this fixed value is
integrated. Similarly, the ghost fields on this line are fixed
at zero.
For definiteness, let us discuss all of the necessary factors
for the vertices in the context of the one loop correction to
the propagator, shown in Fig.~\ref{oneloopws}.
Applying the above considerations to this diagram, we have,
omitting for compactness the delta functions that implement
the Dirichlet conditions on the boundary $i=M$, 
\begin{eqnarray}
T^{\rm one loop}_{fi}
&=&\int\prod_{j=1}^N\prod_{i=1}^{M-1}{dc_i^jdb_i^j\over2\pi}d^2q_i^j
{g^2a\over(32\pi^2m)}\prod_{j=k+2}^{k+l}\left[\delta({\bf q}_{M_1}^j
-{\bf q}_{M_1}^{j-1})2\pi\delta(b_{M_1}^j)\delta(c_{M_1}^j)\right]
\nonumber\\
&&\exp\left\{{a\over m}\sum_{j=1}^N\sum_{i=0}^{M-1}
(b_{i+1}^j-b_{i}^j)(c_{i+1}^j-c_{i}^j)
-{a\over2m}\sum_{j=1}^N\sum_{i=0}^{M-1}
({\bf q}_{i+1}^j-{\bf q}_{i}^j)^2\right\}\\
&&\hskip-15pt\exp\left\{-{a\over m}\left[(b_{M_1}^{k+1}-b_{M_1-1}^{k+1})
(c_{M_1}^{k+1}-c_{M_1-1}^{k+1})
+(b_{M_1+1}^{k+1}-b_{M_1}^{k+1})
(c_{M_1+1}^{k+1}-c_{M_1}^{k+1})-b_{M_1}^{k+1}
c_{M_1}^{k+1}\right]\right\}\nonumber\\
&&\exp\left\{-{a\over m}\left[(b_{M_1}^{k+l+1}-b_{M_1-1}^{k+l+1})
(c_{M_1}^{k+l+1}-c_{M_1-1}^{k+l+1})\nonumber\right.\right.\\
&&\hskip3cm\left.\phantom{a\over m}\left.+(b_{M_1+1}^{k+l+1}-b_{M_1}^{k+l+1})
(c_{M_1+1}^{k+l+1}-c_{M_1}^{k+l+1})-b_{M_1}^{k+l+1}
c_{M_1}^{k+l+1}\right]\right\}\nonumber
\end{eqnarray}
The next step is to try to write a formula that systematically
sums over all the planar diagrams. This can be first done on the
lattice we have constructed. The general planar diagram has
an arbitrary number of vertical solid lines. An interior link $j$ of
a solid line at $i=M_1$ is represented by the product of delta functions
\begin{eqnarray}
\delta({\bf q}_{i}^j
-{\bf q}_{i}^{j-1})2\pi\delta(b_{i}^j)\delta(c_{i}^j)
=\int {d{\bf y}_i^j\over2\pi} e^{i{\bf y}_i^j\cdot({\bf q}_{i}^j
-{\bf q}_{i}^{j-1})}\int d{\bar c}_i^jd{\bar b}_i^je^{{\bar b}_i^jb_i^j
+{\bar c}_i^jc_i^j}.
\end{eqnarray}
A simple way to supply such factors is to assign an Ising spin $s_i^j
=\pm1$ to each site of the lattice. We assign $+1$ if the site $(i,j)$ 
is crossed by a vertical solid line, $-1$ otherwise. 
Then we can represent both kinds of link, solid line and dotted
line by the unified factor
\begin{eqnarray}
\int {d{\bf y}_i^j\over2\pi}\int d{\bar c}_i^jd{\bar b}_i^j 
\exp\left\{[i{\bf y}_i^j\cdot({\bf q}_{i}^j-{\bf q}_{i}^{j-1})
+{\bar b}_i^jb_i^j+{\bar c}_i^jc_i^j]P_i^jP_i^{j-1}
+2\pi{\bar b}_i^j{\bar c}_i^j(1-P_i^jP_i^{j-1})/V\right\},
\end{eqnarray}
where we have defined the projector $P_i^j=(1+s_i^j)/2$
and where $V$ is the volume of transverse space.
Associated with the endpoints of each solid line we have to
supply the missing link factors. These factors occur when
$s_i^j=-s_i^{j-1}$. So in the exponent we 
multiply the factors by $(1-s_i^js_i^{j-1})/2$.
Our final formula for the sum of all planar diagrams is therefore
\begin{eqnarray}
T_{fi}&=&
\sum_{s_i^j=\pm1}\int\prod_{j=1}^N
\prod_{i=1}^{M-1}{dc_i^jdb_i^j\over2\pi}d{\bar c}_i^jd{\bar b}_i^j
{d{\bf y}_i^jd{\bf q}_i^j\over2\pi}
\nonumber\\&&
\exp\left\{{a\over m}\sum_{j=1}^N\sum_{i=0}^{M-1}\left[
{(b_{i+1}^j-b_{i}^j)(c_{i+1}^j-c_{i}^j)}
-{1\over2}({\bf q}_{i+1}^j-{\bf q}_{i}^j)^2\right]\right\}\nonumber\\
&&
\exp\left\{\sum_{j=1}^{N}
\sum_{i=1}^{M-1}\left\{[i{\bf y}_i^j\cdot({\bf q}_{i}^j-{\bf q}_{i}^{j-1})
+{\bar b}_i^jb_i^j+{\bar c}_i^jc_i^j]P_i^jP_i^{j-1}
+{2\pi\over V}{\bar b}_i^j{\bar c}_i^j(1-P_i^jP_i^{j-1}
)\right\}\right\}
\label{isingsum}\\
&&\hskip-.8cm\exp\left\{-{a\over m}
\sum_{j=1}^{N}\sum_{i=1}^{M-1}\left[(b_{i}^{j}-b_{i-1}^{j})
(c_{i}^{j}-c_{i-1}^{j})+
(b_{i+1}^{j}-b_{i}^{j})
(c_{i+1}^{j}-c_{i}^{j})-{g\over4\pi}\sqrt{a\over2m}
b_{i}^{j}c_{i}^{j}\right]
{1-s_i^js_i^{j-1}\over2}\right\}\nonumber
\end{eqnarray}
The first exponent in this formula is just the action for the
free propagator. The second exponent takes care of the 
delta function insertions required for each solid line. The final
exponent removes the appropriate ghost links at the
beginning and end of each solid line. Notice that when $g=0$,
this exponential forces $T_{fi}$ to vanish
unless $s_i^j=s_i^{j-1}$ for all $i,j$,
because there would then not be enough ghost factors to
saturate the ghost integrals. Thus solid lines are
eternal if $g=0$, corresponding to the free field case.

We remark that the expression (\ref{isingsum}) sums all the
planar multiloop corrections to the propagator of the
matrix scalar field. The evolving system is therefore in the
adjoint representation of the color group. That is, we have tacitly
assumed that the only solid lines initially and finally are
those at the boundaries of the strip. More general initial and
final states are described by allowing more solid lines initially
and finally. When the system is in a color singlet state,
we must of course include diagrams in which the outer boundaries
are identified, \ie\ the diagrams should be drawn on a
cylinder, not a strip. In this case, strict periodicity
$q(0)=q(p^+)$ is only possible in the state of zero
total transverse momentum. 

Finally we comment on the possibility that the sum of
planar diagrams (\ref{isingsum}) might describe some
kind of string theory. In finite order in perturbation theory
there are only a finite number of temporal links, and
the worldsheet theory stays topological. However, the 
complete sum might describe a condensation of many temporal
links, providing a bulk dynamics and a possible string
interpretation. In terms of the original
Feynman graphs, this would correspond to the dominance
of the large fishnet diagrams.
Such a condensation would be reflected by
a non zero expectation value of $(1-s_i^js_i^{j-1})$,
or in other words an anti-ferromagnetic ordering of the
Ising spins. Whether or not this happens in a 
given quantum field theory is a detailed dynamical
issue beyond the scope of this article. However,
we hope that the worldsheet description we have
developed will be the right framework for settling
the matter.

\section{Ising spins to Grassmann Integrals: the continuum limit}
\vskip 9pt

Eq.~(\ref{isingsum}) sums perturbation theory in the light cone frame, using the path
integral approach and lattice regularization. Our goal is to derive the
underlying field theory on the world sheet by going to the continuum
limit. The spin variables $s_{i}^{j}$ are well suited to the lattice
description, but they are not convenient for taking the field theory
continuum limit. In this section, we will show how to trade the spin
variables for a pair of anti-commuting variables, suitable for the
field theory description. Rather than reproduce eq.(\ref{isingsum}) in detail in
this new language, we will first explain the general idea in terms of
a simplified version of the model, and then we will proceed directly
to the continuum limit. The key idea is the transfer matrix
$T^{j+1,j}$, which maps from states at $\tau=mj$ to $\tau=m(j+1)$, so that
the $\tau$ evolution of the system can be generated by repeated
applications of the transfer matrix. For example, the bosonic part
of the transfer matrix at site $i$ with ``spin up'',
 corresponding to eq.(\ref{isingsum}), neglecting the
ghosts, is, 
\be
T^{j+1,j}= \exp\left( i{\bf y}^{j+1}\cdot({\bf q}^{j+1}-{\bf q}^{j})\right),
\ee
where we have also suppressed the index $i$ and integration over ${\bf y}$
and ${\bf q}$ for simplicity. Now we introduce a pair of anticommuting
variables $e^{j}$ and $\bar{e}^{j}$ and write the following identities
\bq
\int d\bar{e}^{j} de^{j} \exp(-\bar{e}^{j} e^{j+1} T^{j+1,j}
+ \bar{e}^{j} e^{j})\, e^{j}&=& T^{j+1,j} e^{j+1}\nonumber\\
\int d\bar{e}^{j} de^{j} \exp(-\bar{e}^{j} e^{j+1} T^{j+1,j}
+\bar{e}^{j} e^{j})\, 1 &=& 1.
\eq
 
Here, the factor $e^{j}$ represents an initial state at site $j$
 with spin up and
$1$ an initial state with spin down, and the fermionic integral
correctly maps them into the states at site $j+1$, multiplied by
the transfer matrix (the transfer matrix for spin down state is unity).
Iterating this yields the transfer matrix between the initial and
the final states:
\be
T_{fi}=\int \prod_{j} d\bar{e}^{j} de^{j} \exp\left(
\sum_{j}(-\bar{e}^{j} e^{j+1} T^{j+1,j}+ \bar{e}^{j} e{j}\right).
\ee
The continuum limit is now easy. As we let the lattice spacing
$a\rightarrow 0$
$$
e^{j+1}-e^{j}\rightarrow a \dot{e}(\tau),\;T^{j+1,j}\rightarrow
1+ i\, a\, H(\tau),
$$
and,
\be
T_{fi}\rightarrow \int De D\bar{e}
\, \exp\left(\int d\tau (- \bar{e}(\tau)
\dot{e}(\tau)- i \bar{e}(\tau) e(\tau) H(\tau))\right),
\ee
where differentiation with respect to $\tau$ is designated by a dot.
 In the case considered above,
$H= {\bf y}\cdot\dot{{\bf q}}$.

The construction given above does not allow the flipping of the spin.
 This is because the spin up and spin down states have opposite fermionic
 gradings, and the transfer matrix, having even grading, cannot connect
one state to the other. This problem can be overcome by doubling the
number of anticommuting variables. Labeling them by indices 1 and 2,
$e_{1,2}$ inserted in the path 
integral correspond to spin up and spin down states respectively.
$T$ is now given by
\bq
T_{fi}&=&\int De_{1} D\bar{e}_{1}
 De_{2} D\bar{e}_{2}
 \exp\Big(\int d \tau
(-\bar{ e}_{1} \dot{e}_{1}
-\bar{e}_{2} \dot{e}_{2}- i\bar{e}_{1} e_{1} H\nonumber\\
 &&\hskip1cm +g V \bar{e}_{1} e_{2} +g V \bar{e}_{2} e_{1})\Big).
\eq
The terms multiplied by the coupling constant g and the interaction
vertex $V$ are the 
spin flipping terms. They correspond to beginning or ending of a
solid line, where the interaction takes place.

The model presented above is an oversimplified version of the realistic
model, which includes fermionic ghosts and special vertices at the
points of interaction, which will be specified later. Adding the ghosts and
restoring the index $i$, or rather its continuum version $i m\rightarrow 
\sigma$, which has been suppressed so far, we have
\bq
T_{fi}&=&\int D{\bf q}\,D{\bf y}\, Db\, D\bar{b}
\, Dc\, D\bar{c}\, De_{1}\, D\bar{e}_{1} 
\,De_{2}\, D\bar{e}_{2}\nonumber\\
& &\times \exp\Big(\int d\sigma d\tau \Big[- \frac{1}{2}
\left({\bf q}'\right)^{2}+ b'c'+ i\bar{e}_1 e_1 (
{\bf y}\cdot\dot{{\bf q}}+ \bar{b} b+ \bar{c} c)\nonumber\\
& &+ \bar{e}_{1}\dot{e}_{1}+\bar{e}_{2}\dot{e}_{2}
+ g V\,\bar{e}_{1}e_{2}+ g  V\,\bar{e}_{2}e_{1}\Big]\Big).
\label{grasseq}
\eq

It remains to specify the vertex $V$. The lattice version of
this vertex is the factor in eq.(\ref{isingsum})
\be
V^{(i,j)}=
\exp\left\{-\frac{a}{m}\left[(b_{i}^{j}-b_{i-1}^{j})(c_{i}^{j}-
c_{i-1}^{j})+ (b_{i+1}^{j}-b_{i}^{j})(c_{i+1}^{j}-c_{i}^{j})
\right]\right\}
\ee
to be inserted at the site labeled by $(i,j)$,
 the start or at the end of a solid line, as shown
in the figure. This can be replaced by an equivalent vertex
which is more suitable for taking the continuum limit:
\be
V^{(i,j)}\rightarrow \exp\left(-\frac{a}{m}(b_{i-1}^{j}
c_{i-1}^{j}+ b_{i+1}^{j} c_{i+1}^{j})\right).
\ee
Before taking the continuum limit, we will expand the exponential
in a power series. This series has only four terms,
\be
V^{(i,j)}= 1- \frac{a}{m}(b_{i-1}^{j} c_{i-1}^{j} +
b_{i+1}^{j} c_{i+1}^{j})+ \frac{a^{2}}{m^{2}}( b_{i-1}^{j}
c_{i-1}^{j} b_{i+1}^{j} c_{i+1}^{j}).
\ee
The first term in the series is the classical contribution; namely,
the solid lines beginning or ending without any insertions. In the
lattice version, the remaining terms were needed to get the correct
factors of $p^{+}$ multiplying the propagators. We consider these as
quantum corrections to the classical result. To have a sensible
continuum limit for the first term, we should scale $g$ as $
g=a^{2} g_{1},$
where $g_{1}$ is finite in the limit the lattice spacings $a$ and $m$
go to zero. The second and the third terms are multiplied by the
factor $a/m$, which is ambigious in the continuum limit. We will
provisionally assign an independent coupling constant $g_{2}$ to
these terms. The final term formally vanishes upon identifying
the index $i-1$ with $i+1$ in the continuum limit. There remains,
however, a residual term if we expand to second order in the lattice
spacing $m$. Assigning another coupling constant $g_{3}$ to this
term, the last two terms in eq.(40) can now be written as
$$
g V (\bar{e}_{1} e_{2}+ \bar{e}_{2} e_{1}) \rightarrow
(\bar{e}_{1} e_{2}+\bar{e}_{2} e_{1})(g_{1} + g_{2} b c
+ g_{3} b' c').
$$
This completes the discussion of the continuum world sheet action.
It is somewhat formal, since we have not investigated the question
of renormalization. For example, Lorentz invariance should
imply certain relations between the constants $g_{1}$, $g_{2}$
and $g_{3}$, which can only be uncovered by a careful study of
renormalization. We hope to address these questions in the future.

\section{Mass Terms}
So far we have dealt exclusively with the massless propagator
in $\Tr\phi^3$ field theory. Including a mass for the field
is not completely trivial, because the mixed
representation of the scalar field propagator would acquire the
extra factor $e^{-a\mu^2/2lm}$, which is not local on the
world sheet. On the other hand, we know that in the cubic
theory, mass renormalization is unavoidable, so we {\it must
at least} be able to insert mass counter-terms. 

\begin{figure}[ht]
\psfrag{'k'}{$k$}
\psfrag{'k+1'}{$k+1$}
\psfrag{'M'}{$M$}\psfrag{'M1'}{$M_1$}
\psfrag{'N'}{$N$}
\centerline{\epsfig{file=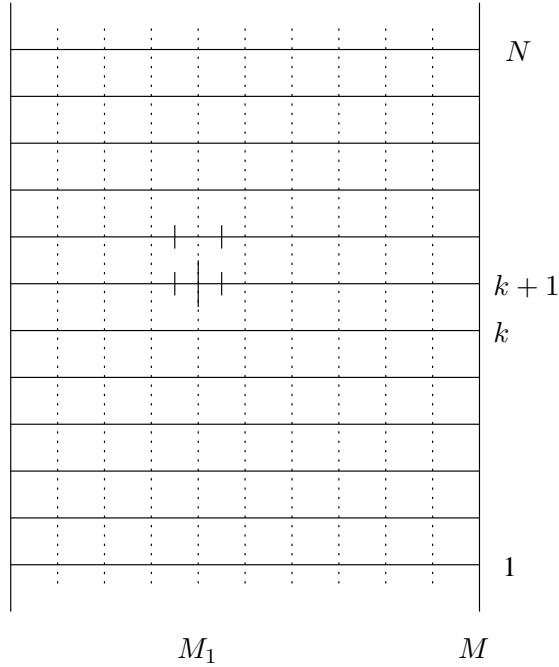,width=8cm}}
\caption{A mass insertion represented as a ``short loop''.}
\label{massinsert}
\end{figure}

Actually, this latter observation
points to one way of introducing mass, because the self
energy divergence arises precisely in the short time
limit of the virtual loop. Thus we can introduce
a mass insertion by a short loop. For example, we
could take it to be the $l=1$ case of the loop
shown in Fig.~\ref{oneloopws}. This is shown explicitly
in Fig.~\ref{massinsert}. 
Note that a cluster of four
ghost links is missing in this diagram, and there
are no extra momentum conserving delta functions.
Thus the insertion will provide a factor of $1/M^2$.
If we remember that on a fixed time slice the
insertion can be placed on $M-1$ sites, we see that
the net factor is $(M-1)/M^2\sim 1/M$ for the 
continuum limit $M\to\infty$. This is precisely
the $M$ dependence required of a perturbative mass
insertion: $e^{-a\mu^2/2Mm}\sim 1-(a/m)\mu^2/2M$.
In the Ising spin representation of the sum
over diagrams this mass insertion would be multiplied
by the Ising spin factors $(1-s_i^{j-1})(1+s_i^{j})
(1-s_i^{j+1})/8$:
\begin{eqnarray}
{\rm Mass~Term}&=&-{a\over m}
\sum_{i,j}\left[(b_i^j-b_{i-1}^j)(c_i^j-c_{i-1}^j)
+(b_{i+1}^j-b_{i}^j)(c_{i+1}^j-c_{i}^j)+\mu\sqrt{a\over2m}b_i^jc_i^j
\right.\nonumber\\
&&\quad\left.+(b_i^{j+1}-b_{i-1}^{j+1})(c_i^{j+1}-c_{i-1}^{j+1})
+(b_{i+1}^{j+1}-b_{i}^{j+1})(c_{i+1}^{j+1}-c_{i}^{j+1})
-\mu\sqrt{a\over2m}b_i^{j+1}c_i^{j+1}\right]\nonumber\\
&&\hskip4cm \times{(1-s_i^{j-1})(1+s_i^{j})
(1-s_i^{j+1})\over 8}.
\end{eqnarray}
Note that since we want the mass insertion to give a negative
prefactor $-\mu^2$ we have used opposite signs in the two
terms proportional to $\mu$.

If there are several fields with different mass, introducing
a mass insertion into the bulk in this way would not
be a local procedure since a whole strip must then change
its mass simultaneously. Indeed, even if the mass were the
same for two fields of different ``flavor'' it is
clear that the flavor information must reside on the
boundaries in order for flavor changing transitions
to be local on the worldsheet. Thus in this case
flavor dependent mass differences must also reside
on the boundaries. This is certainly feasible: 
just use an insertion in which there is only one
missing ghost link next to the boundary. This
would provide a factor of $1/M$, which is correct,
since there are not order $O(M)$ places to insert
on a fixed time slice. 

\section{Dualities}
It is interesting and amusing to compare the lattice
world sheet we have set up to sum planar QFT with
the lattice string formalism set up in \cite{gilest}
to sum planar open-string multi-loop diagrams.
The two pictures are dual in almost every
conceivable meaning carried by that word.

First of all the lattice string formalism uses
transverse coordinates as basic variables
while the QFT world sheet uses transverse momentum
variables. Secondly, the basic variables satisfy
Neumann boundary conditions in the lattice
string theory but Dirichlet boundary
conditions in the QFT. 
Thirdly, a loop in the lattice string
formalism is described by a row of missing spatial
links, whereas a loop on the QFT worldsheet is
represented by a row of {\it added} temporal
links. Finally the role of strong and weak
coupling is reversed in the two pictures:
at zero coupling every spatial link is present
on the string worldsheet, but at zero coupling
on the QFT worldsheet no temporal link is
present. Conversely, at strong coupling a maximum
number of spatial links are missing
on the string world sheet, whereas a maximal
number of temporal links are present on the
QFT worldsheet. 

\section{Directions for Future Work}
In this paper we have concentrated on the task of developing
a world sheet description for the simplest possible matrix
quantum field theory: a scalar theory with cubic interactions.
When we turn to richer theories there are a number of
technical problems that emerge in the application of
our methods.

Consider first the simple complication of including
several types (flavors) of field. If different fields
couple to each other, a local worldsheet description
requires that the ``bits'' of the different fields
are the same ``stuff''. Whatever differences there
are must occur at the boundaries. This suggests
a Chan-Paton factor approach to flavor
as in open string theory.

When we move on to consider fields with spin, life
gets more interesting. Do we put the spin variables
only on the boundaries or can we give the bits
spin degrees of freedom in such a way that 
the total spin carried by the strip is fixed
by boundary values of these variables? At
the moment both seem viable possibilities,
but we do not delve into them here.

Finally, what about quartic interactions? Our
methods obviously work well with cubic vertices,
but how can we deal in a worldsheet local way with
quartic vertices? One way, but hopefully not the
only way, is to represent quartic vertices as
a concatenation of a pair of cubic vertices,
as was done for example in \cite{beringrt}.
If this is to work the intermediate ``pseudoparticle'' that
connects the two cubic vertices should
not propagate very long. This can be 
arranged, for example, by giving the intermediary a huge
mass. Such a scheme can be made to work, but
we hope a more elegant approach will emerge from
future research.

It is a clear challenge to bring all interesting planar
quantum field theories into a worldsheet
description along the lines proposed in this article.
Much work still needs to be done.

\vskip.5cm
\noindent\underline{Acknowledgments:}
This work was supported in
part by the Department of Energy under Grants No. DE-FG02-97ER-41029
and DE-AC03-76SF00098, and in part by tne National Science Foundation
Grant PHY-0098840. Also
CBT acknowledges support from the Miller Institute for Basic
Research in Science.

\bibliography{../larefs}
\bibliographystyle{unsrt}

\end{document}